**Introduction: Toward an interdisciplinary science of spontaneous thought**

Kieran C. R. Fox[a] and Kalina Christoff[b,c]

[a] Department of Neurology and Neurological Sciences, Stanford University, Stanford, CA, U.S.A., 94304

[b] Department of Psychology, University of British Columbia, Vancouver, B.C., Canada, V6T 1Z4

[c] Centre for Brain Health, University of British Columbia, 2211 Wesbrook Mall, Vancouver, B.C., V6T 2B5 Canada

Where do spontaneous thoughts come from?

It may be surprising that the seemingly straightforward answers "from the mind" or "from the brain" are in fact an incredibly recent, modern understanding of the origins of spontaneous thought. For nearly all of human history, our thoughts – especially the most sudden, insightful, and important – were almost universally ascribed to divine or other external sources. Cultures around the world believed that dreams were messages sent from the gods (Kracke, 1991); inventions like writing and agriculture were credited to ancient culture heroes and tutelary deities long lost in the mists of legend (Chang, 1983); and the belief that artistic creativity was inspired by the Muses (Murray, 1989) held sway for two millennia (McMahon, 2013). Even the original sense of the word *inspiration* was that the divine had been 'breathed into' a mere mortal, accounting for the new idea or insight. There were of course exceptions – Aristotle, for instance, put forward the naturalistic hypothesis that dreams were created by the mind of the dreamer (Aristoteles & Gallop, 1996) – but nowhere, it seems, was there a widespread belief in the spontaneity, originality, and creativity of the unaided human mind.

We still sometimes worship our great intellectual innovators – artists, scientists, philosophers – as semi-divine figures. But somewhere, somehow, our perspective changed and we began to see ourselves as the authors of our own thoughts, however inexplicable their origins might seem. Perhaps the beginnings of this shift in perspective are echoed in the ancient myth of Prometheus, who "stole and gave to mortals" the "fount of the arts, the light of fire" – in other words, the power of conjuring up novel thoughts (Griffith, 1983). But although this internalization of thought's origins began long ago, only in the past few centuries have human beings truly taken responsibility for their own mental content, and finally localized thought to the central nervous system – laying the foundations for a protoscience of spontaneous thought.

This shift has broadly answered the *who* and the *where* of spontaneous thought: we are the source of our thoughts, and these thoughts seem to be constructed in our heads. But enormous questions still loom: *what*, exactly, is spontaneous thought? *Why* does our brain engage in spontaneous forms of thinking, and *when* is this most likely to occur? And perhaps the question most interesting and accessible from a scientific perspective: *how* does the brain generate, elaborate, and evaluate its own spontaneous creations? Each chapter that follows aims to provide at least preliminary answers to these perplexing questions.

The central aim of this volume is to bring together views from neuroscience, psychology, philosophy, phenomenology, history, education, contemplative traditions, and clinical practice in order to begin to address the ubiquitous but poorly understood mental phenomena that we collectively call 'spontaneous thought.' Perhaps no other mental experience is so familiar to us in daily life, and yet so difficult to understand and explain scientifically. The present volume represents the first effort to bring such highly diverse perspectives to bear on answering the what, when, why, and how of spontaneous thought.

Although 'spontaneous thought' as a term has been used throughout the last decade in both the psychological (Klinger, 2008) and neuroscientific literature (Christoff, Irving, Fox, Spreng, & Andrews-Hanna, 2016; Christoff, Ream, & Gabrieli, 2004; Fox, Spreng, Ellamil, Andrews-Hanna, & Christoff, 2015), recent years have marked tremendous progress in our theoretical understanding of what spontaneous thought is and what phenomena it encompasses. Spontaneous thought can be defined as thought that arises relatively freely due to an absence of strong constraints on its contents or on the transitions from one mental state to another (Christoff et al., 2016). In other words, spontaneous thought moves freely as it unfolds.

*Figure 1.* Conceptual space relating different types of thought and their constraints.

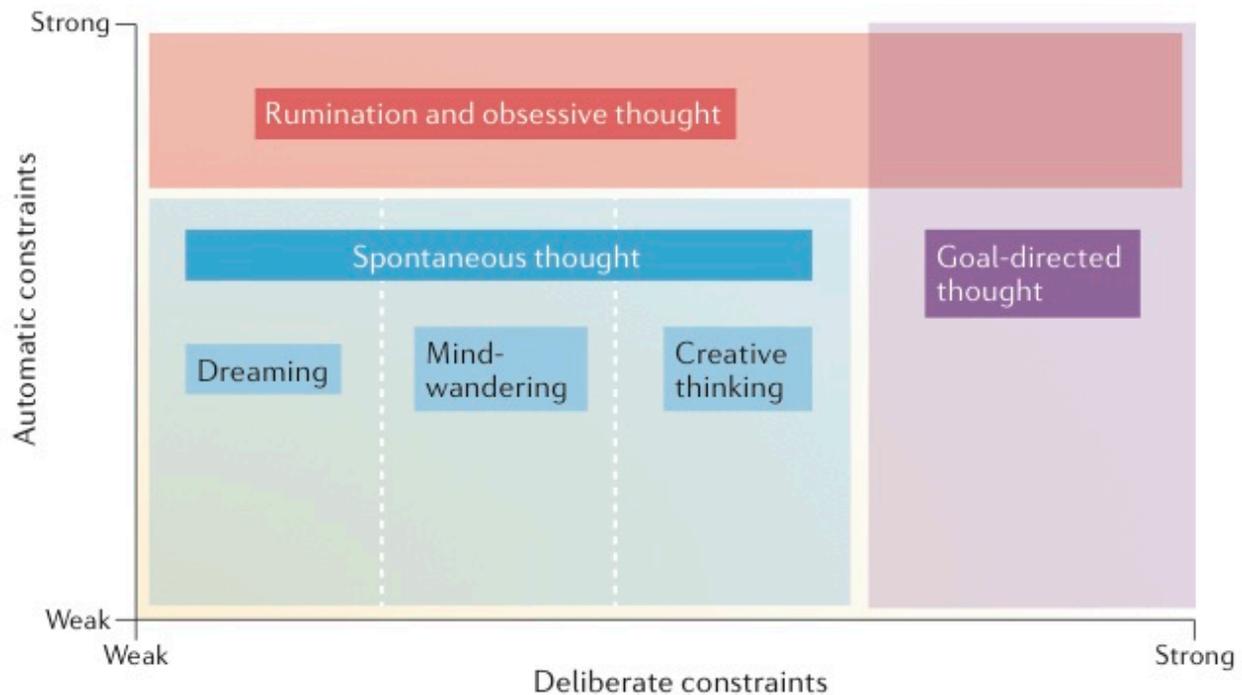

Reproduced, with permission, from Christoff et al. (2016).

There are two general ways in which thought can be constrained (Fig. 1). One type of constraint is flexible and deliberate, and implemented through cognitive control. Another type of constraint is automatic in nature. Automatic constraints can be thought of as a family of mechanisms that operate outside of cognitive control to hold attention on a restricted set of information. Examples of automatic constraints are emotional significance and habits, both of which can constrain our thoughts without any effort or intention on our part (Christoff et al., 2016; Fox, Kang, Lifshitz, & Christoff, 2016; Todd, Cunningham, Anderson, & Thompson, 2012).

Spontaneous thought can also be understood as a broader family of mental phenomena, including our daytime fantasies and mind-wandering (Christoff, 2012; Fox et al., 2015); the flashes of insight and inspiration familiar to the artist, scientist, and inventor (Beaty, Benedek, Silvia, & Schacter, 2015; Kounios & Beeman, 2014; Zabelina & Andrews-Hanna, 2016); and the nighttime visions we call dreams (Christoff et al., 2016; Domhoff & Fox, 2015; Fox, Nijeboer, Solomonova, Domhoff, & Christoff, 2013). There is a dark side to spontaneous thought as well – the illumination of which is yet another major goal of this volume. Repetitive depressive rumination; uncontrollable thoughts in obsessive-compulsive disorder; the involuntary and lifelike re-experiencing of post-traumatic stress disorder – all these, we suggest, can be considered dysfunctional forms of spontaneous thought, and need to be understood in relation to our natural and healthy propensity toward novel, variable, imaginative thought ((Christoff et al., 2016); see also Mills et al., this volume).

*Spontaneous* should in no way suggest *random* or *meaningless*. Another key aim of this volume is to highlight the ample evidence in favor of the idea that goal-related and 'top-down' processing often co-occurs with and can sometimes guide spontaneous thought (Fox & Christoff, 2014; Fox et al., 2016; Klinger & Cox, 2011; Seli, Risko, Smilek, & Schacter, 2016). Although the cause and meaning of specific thoughts or dreams often eludes us, the rare but sensational occurrences of transgressive thoughts or highly bizarre and emotional dreams tend to obscure just how mundane (but quite possibly, useful) most of our self-generated mental content really is (Domhoff, 2003; Fox et al., 2013; Klinger, 2008). The degree to which mental processes that are ostensibly spontaneous and beyond our control appear to be planned, relevant, and insightful with respect to our personal goals and concerns is striking – and, we believe, deserving of further exploration.

These ubiquitous spontaneous mental phenomena raise some intriguing questions: Can we engage in planning and other executive processes in the absence of conscious awareness? To what extent are 'we' in control of our own minds? The true qualities and content of spontaneous thought also fly in the face of many culturally sanctioned but unwarranted beliefs about the inexplicability of our fantasy lives, the randomness and meaninglessness of dreams, or the disorderliness of creative thoughts and insights in artists and scientists. A closer look at psychological, neuroscientific, and philosophical work shows not only the co-occurrence of cognitive processes like planning, mentalizing, and metacognition with various forms of spontaneous thought (Christoff, Gordon, Smallwood, Smith, & Schooler, 2009; Fox & Christoff, 2014; Fox et al., 2015; Seli et al., 2016), but also a compelling correspondence between the content of one's spontaneous thoughts and the content and concerns of one's daily life (Klinger, 2008). *The Oxford Handbook of Spontaneous Thought* is the first volume of its kind to bring together experts from so many diverse fields to explore these phenomena, and should therefore be of interest to psychologists, neuroscientists, philosophers, clinicians, educators, and artists alike; indeed, to anyone intrigued by the incredibly rich life of the mind.

*Overview of the Handbook*

This Handbook is divided into seven separate but closely interrelated parts. This introductory chapter comprises Part One, providing an overview of spontaneous thought in general and the many chapters of this book in particular.

Part Two dives right into fundamental theoretical issues surrounding the definition and investigation of spontaneous thought. Caitlin Mills, Arianne Herrera-Bennett, Myrthe Faber, and Kalina Christoff ask why the mind wanders at all, and propose the Default Variability Hypothesis: the idea that by default, spontaneous thought tends to exhibit high variability of content over time – variability that serves as an adaptive mechanism that enhances episodic memory efficiency and facilitates semantic knowledge optimization. Chandra Sripada puts forward a theoretical framework within which spontaneous and deliberate thought can be compared, respectively, with 'exploration' of the environment in search of new resources vs. 'exploitation' of the resources we already have at hand. Carey Morewedge and Daniella Kupor provide an overview of people's metacognitive appraisals of the meaning and relevance of spontaneous thoughts, with the surprising conclusion that people tend to attribute *more* importance to thought whose origin is mysterious – perhaps hearkening back to the ancient human view of the origins of thought discussed at the outset of this chapter. Dylan Stan and Kalina Christoff propose that a key quality of mind-wandering is an accompanying subjective experience of *ease*, or low motivational intensity. Georg Northoff proposes a novel theory aiming to explain how spontaneous brain activity generates and constitutes subjectively-experienced spontaneous thought. Finally, Jonathan Smallwood, Daniel Margulies, Boris Bernhardt, and Elizabeth Jeffries present their component process framework of spontaneous thought, explaining how different types of thought can arise through the interaction of specific underlying neurocognitive processes.

Part Three explores the broader philosophical, evolutionary, and historical perspectives on spontaneous thought. Zachary Irving and Evan Thompson provide an in-depth introduction to the philosophy of mind-wandering, reviewing several psychological and philosophical accounts and providing a new view of their own. In his chapter, Thomas Metzinger addresses the question, 'Why is mind-wandering interesting for philosophers?' Dean Simonton relates the spontaneity of human thought to other spontaneous generative processes, highlighting the connections with 'selectionist' views of evolution and creativity. John Antrobus offers an analysis of how the brain in both waking and sleeping can so effortlessly produce a constant stream of visual imagery and thoughts – and what use they might have. Rounding out the section, Alex Soojung-Kim Pang explores how spontaneous thought was viewed in the past and used by creative people to further their endeavors, and how deep historical research could lead to an understanding of the role of spontaneous thought in the history of ideas.

Part Four focuses on mind-wandering and daydreaming. In the first chapter, Jessica Andrews-Hanna, Zachary Irving, Kieran Fox, Nathan Spreng, and Kalina Christoff present an interdisciplinary overview of the rapidly-evolving neuroscience of spontaneous thought. Investigating what we have learned from intracranial electrophysiology in humans, Kieran Fox then synthesizes the available evidence on how and where self-generated thought is initiated within the brain. Arnaud D'Argembeau provides a detailed discussion of the link between mind-wandering and self-referential thinking, and their common neural basis. David Stawarczyk provides a detailed overview of the phenomenological properties of all kinds of mind-wandering and daydreaming, covering both the historical trajectory of these investigations and the present state of research. Eric Klinger, Igor Marchetti, and Ernst Koster discuss how critical goal-pursuit is to spontaneous thought, elaborating on how these thoughts are adaptive in everyday life but can go awry in a variety of clinical conditions. Claire Zedelius and Jonathan Schooler provide a fine-grained view of the many different kinds of mind-wandering and the evidence that they have distinctive effects on task performance, mood, and creativity. Julia Kam and Todd Handy comprehensively review the evidence from human electrophysiology that mind-wandering involves a decoupling of attention from the external world. Finally, Jeffrey Wammes, Paul Seli, and Daniel Smilek review what we know about mind-wandering in educational settings, and how excessive, unintentional mind-wandering in the classroom impacts learning and academic performance.

Part Five covers creativity and insight, and their relation to other forms of spontaneous thought. Roger Beaty and Rex Jung offer an overview of how large-scale brain networks interact during creative thinking and creative performance. Mathias Benedek and Emanuel Jauk offer detailed empirical evidence for a 'dual-process' model of creative cognition, wherein the flexible switching between controlled and spontaneous cognition is critical to an optimal creative process. Charles Dobson, himself an artist as well as professor of fine arts, offers an insider's view of what he calls 'flip-flop thinking', and outlines his first-hand experiences of what helps (and what hurts) the creative process. John Vervaeke, Leo Ferraro, and Arianne Herrera-Bennett develop an intriguing account of the 'flow' state as a form of spontaneous thought characterized by a cascade of successive insights and learning experiences. Oshin Vartanian delves into how self-referential thoughts can be elicited by aesthetic appreciation of artworks, such as paintings. Finally, David Beversdorf provides an extensive review of the neurochemical basis of flexible and creative thinking.

Spontaneous thought does not cease when we close our eyes and turn out the lights. Part Six explores the many normal, extraordinary, and sometimes pathological varieties of spontaneous thought that take place throughout the sleep cycle, and how these are related to memory consolidation and involuntary memory retrieval. G. William Domhoff provides an overview of the neural basis of dreaming and REM sleep, while the chapter by Kieran Fox and Manesh Girn provides a comprehensive review of what is known about the neural correlates of all sleep stages throughout the sleep cycle. Jennifer Windt and Ursula Voss provide an in-depth treatment of the phenomenon of lucid dreaming, bringing psychological, philosophical, and neuroscientific evidence to bear to better explain this remarkable mental state. Tore Nielsen explores the fascinating topic of 'microdreaming' and hypnagogic imagery as a paradigm for a fine-scaled neurophenomenological approach to inner experience. Elizaveta Solomonova offers an interdisciplinary look at the little-known phenomenon of sleep paralysis, and the spontaneous visions and emotions that accompany it. Erin Wamsley explores how spontaneous thought in both waking and sleep can be seen as an expression of memory consolidation and recombination, and John Mace provides a comprehensive overview of involuntary memories – how often they occur, how they can chain together, how they differ from voluntarily recalled information, and what their function might be.

Part Seven takes us to the fringes and also the cutting edge of research on spontaneous thought: its relationship to clinical conditions and altered states of consciousness. Dylan Stan and Kalina Christoff begin the section by outlining the many potential clinical benefits and risks of spontaneous thought. Claire O'Callaghan and Muireann Irish describe the neural underpinnings of how spontaneous thought changes in relation to aging and dementia syndromes. Elizabeth DuPre and Nathan Spreng explore the relationships between depression, rumination, and spontaneous thought. Aaron Kucyi explores the intriguing relationships between mind-wandering and both chronic and acute pain, and how these interactions are mediated by large-scale brain networks. Halvor Eifring investigates how religious and contemplative traditions around the world have tended to see mind-wandering as an obstacle, while at the same viewing spiritual attainment and liberation as a spontaneous process of transformation that cannot be actuated deliberately. Wendy Hasenkamp outlines how meditation and mindfulness practices can provide a window into the rapid fluctuations of mind-wandering. Peter Suedfeld, Dennis Rank, and Marek Malůš offer an account of spontaneous thought in extreme and unusual environments, exploring rarely-seen records of the thoughts and experiences of polar explorers, astronauts, and those undergoing severe sensory deprivation. Last but by no means least, Michael Lifshitz, Eli Sheiner, and Laurence Kirmayer detail how the powerful unconstrained cognition brought on by psychedelic substances can be guided by culture and context.

    All told, these 40 chapters provide the most comprehensive overview of the wide-ranging field of spontaneous thought to date – and there could be no better guides to this realm than the 64 outstanding scientists, historians, philosophers, and artists who have come together to write them.

Spontaneous forms of thought enable us to transcend not only the here and now of perceptual experience, but also the bonds of our deliberately-controlled and goal-directed cognition; they allow the space for us to be other than who we are, and for our minds to think beyond the limitations of our current viewpoints and beliefs. In studying such an abstruse and seemingly impractical subject, we need always to remember that our capacity for spontaneity, originality, and creativity defines us as a species – and as individuals.

The painting adorning the cover of this Handbook is by artist and neuroscientist Greg Dunn, who draws inspiration for his work from the ancient s*umi-e* tradition of ink wash painting still practiced in Japan. The essence of *sumi-e*, which has deep roots in Taoism and Zen Buddhism, is to combine discipline with spontaneity, to evoke a complex essence with simplicity; to bring order, so to speak, out of chaos, and give rise to a creation that is coherent and integrated, yet natural and unforced (Cheng, 1994; Van Briessen, 2011; Watts, 1957). We could think of no better artist to provide a visual overture to the multifaceted exploration of these very same themes throughout the pages of this book.

Philosopher Alan Watts eloquently captured the tension and interplay between spontaneity and purpose when he wrote: "spontaneity is not by any means a blind, disorderly urge, a mere power of caprice. A philosophy restricted [by] conventional language has no way of conceiving an intelligence which does not work according to plan, according to a one-at-a-time order of thought. Yet the concrete evidence of such an intelligence is right to hand…" (Watts, 1957). We hope the chapters that follow help to illuminate this elusive wisdom of spontaneous thought in all its many manifestations.

# References

...